\documentclass[12pt]{iopart}

\usepackage{iopams}
\usepackage{amstext}
\usepackage{amssymb} 
\usepackage{latexsym}
\usepackage[english]{babel}
\usepackage[dvips]{graphicx}

\newcommand{\beq}{\begin{equation}}
\newcommand{\eeq}{\end{equation}}
\newcommand{\ket}[1]{\left\vert#1\right\rangle}
\newcommand{\bra}[1]{\left\langle#1\right\vert}
\newcommand{\Ham}{\mathcal H}

\begin{document}

\title[Adiabatic dynamics in a spin-1 chain with uniaxial single-spin anisotropy]
{Adiabatic dynamics in a spin-1 chain with uniaxial single-spin anisotropy}

\author{Elena Canovi$^{1,2}$, Davide Rossini$^1$, Rosario Fazio$^3$ and Giuseppe E. Santoro$^{1,4,5}$}

\address{$^1$ \, International School for Advanced Studies (SISSA),
  Via Beirut 2-4, I-34014 Trieste, Italy}
\address{$^2$ \, Istituto Nazionale di Fisica Nucleare (INFN), Sezione di Trieste,
  Via Valerio 2, I-34127 Trieste, Italy}
\address{$^3$ \, NEST-CNR-INFM \& Scuola Normale Superiore,
  Piazza dei Cavalieri 7, I-56126 Pisa, Italy}
\address{$^4$ \, International Centre for Theoretical Physics (ICTP),
  P.O. Box 586, I-34014 Trieste, Italy}
\address{$^5$ \, CNR-INFM Democritos National Simulation Center,
  Via Beirut 2-4, I-34014 Trieste, Italy}
  
\ead{canovi@sissa.it, drossini@sissa.it, fazio@sns.it and santoro@sissa.it}

\date{\today}

\begin{abstract}

We study the adiabatic quantum dynamics of an anisotropic spin-1 XY chain
across a second order quantum phase transition.
The system is driven out of equilibrium by performing a quench on the
uniaxial single-spin anisotropy, that is supposed to vary linearly in time.
We show that, for sufficiently large system sizes, the excess energy
after the quench admits a non trivial scaling behavior that is not predictable
by standard Kibble-Zurek arguments for isolated critical points or extended
critical regions.
This emerges from a competing effect of many accessible low-lying excited
states, inside the whole continuous line of critical points.

\end{abstract}

\pacs{75.10.Jm, 73.43.Nq, 64.60.Ht}


\vspace*{1.cm}

\noindent{\it Keywords:\/} Spin chains, ladders and planes (theory);  Quantum phase transitions (theory); Density matrix renormalization group calculations

\maketitle

\section{Introduction}

Recent impressive experimental advances in manipulating cold atoms
loaded in optical lattices have opened up the possibility to investigate the
actual dynamics of quantum many-body systems with very low dissipation rates
and long coherence times~\cite{bloch08}; this also allowed a very accurate
check of the fundamental laws describing the physics of such systems.
Among the others, it has been possible to probe a variety of very interesting
and genuinely quantum non-equilibrium phenomena, such as, for example, the 
collapse and the revival of a Bose-Einstein condensate~\cite{greiner02},
the manipulation of the atomic number statistics~\cite{orzel02},
or the coherent non-equilibrium evolution of one-dimensional strongly
interacting bosons from a carefully prepared initial state~\cite{kinoshita06}.
Furthermore, non-equilibrium in cold atomic gases can also be achieved by 
changing in time some of the coupling constants of the system, e.g., the depth of 
the optical lattice or the harmonic trap, on a scale shorter than the relaxation 
rate. These new experimental capabilities have spurred a renewed interest in 
the study of {\em quantum quenches}. 

A lot of attention has been devoted to the study of sudden quenches (see 
for example~\cite{quantumquenches} and references therein). In this paper 
we deal with an equally debated problem, i.e., when the changes in the 
coupling constants driving the quantum system are performed adiabatically. 
This problem becomes non-trivial if, during the quench, the system crosses a 
Quantum Phase Transition (QPT). Due to the closure of the gap in the thermodynamic 
limit, the system will be unable to stay in its equilibrium ground state, no matter how 
slow is the quench. This problem plays a crucial role in adiabatic quantum 
computation schemes, where the system Hamiltonian is supposed to be slowly changed
on a time scale that is large as compared to the typical inverse
zero-temperature gap, so that the system always remains in its instantaneous
ground state~\cite{fahri01,santoro02,santoro06}. The efficiency of an adiabatic quantum computation 
algorithm relies on the assumption that the minimum gap between the ground state and 
the first excited state goes gently to zero in the thermodynamic limit.
When this is not the case, the non-adiabatic evolution close to the QPT drives the system 
out of the ground state. The computation is no-longer accurate or, in other words,
a number of defects appears in the final state.

The problem of defect formation in the adiabatic dynamics of critical systems
was examined much before quantum information:
it was first considered by Kibble and Zurek (KZ) in the context of phase
transitions in the early universe~\cite{kibble,zurek} and more recently extended
to the quantum case~\cite{zurek05,polkov05},
raising an intense theoretical discussion~\cite{damski05,dziarmaga05,damski06,cherng06,
cucchietti07,cincio07,caneva07,caneva08,patane08,pellegrini08,sengupta08,viola08,divakaran08,
sen08,cincio08,gritsev08,degrandi08,schutzhold06}.
According to the KZ mechanism, the evolution of a quantum system
is either adiabatic or impulse, depending on the distance from the critical point.
The time (i.e., the distance from the critical point) at which the system switches from one 
regime to the other depends on the speed of the quench: the slower it is, the later the 
evolution will become impulse. This argument allows to predict the scaling of the density
of defects as a function of the quench rate. Interestingly, for very slow quenches
the quantum evolution can be also successfully studied~\cite{zurek05,damski05} by means 
of an effective two-level approximation with an avoided level crossing,
within the Landau-Zener (LZ) formalism~\cite{landau,zener}.
A more general scenario arises in the presence of non-isolated quantum
critical points, which can accumulate and form extended
critical regions. Here the validity of the KZ mechanism is a priori not obvious,
even if in some cases it is still possible to predict the defect density
by identifying a dominant critical point, or by using scaling 
arguments~\cite{pellegrini08,sengupta08,viola08,divakaran08}.

In this paper we study the adiabatic dynamics in a one-dimensional XY spin-1
system with single-ion uniaxial anisotropy~\cite{schulz86,denijs98},
exhibiting (in equilibrium) a QPT of the Berezinskii-Kosterlitz-Thouless (BKT) type.
Our interest in the dynamics of this specific spin-chain is motivated by the fact
that it describes quite accurately the properties of the Bose-Hubbard (BH) 
Hamiltonian both in the limit of strong interaction and close to the 
Mott-to-superfluid QPT~\cite{altman02,huber07};
understanding the nonlinear response of such system to slow quenches
may reveal itself as a powerful tool to probe Bose condensates
loaded in optical lattices~\cite{degrandi08}.
Interestingly, our results suggest that the knowledge of the
static properties of the system, in particular of the BKT transition,
may not be sufficient to predict and fully characterize the dynamical behavior.

Some dynamical properties of the BH model after a quasi-adiabatic crossing
of the QPT have been analyzed both from the superfluid to the Mott
insulator~\cite{schutzhold06}, and in the opposite
direction~\cite{clark04}, where topological defects arise.
Other works focused on the emergence of universal dynamical scaling,
when quenching to the superfluid phase: they started using
the original KZ mechanism~\cite{polkov05,cucchietti07}, but then realized that,
for non-isolated critical points or critical surfaces,
a generalization in terms of dynamical critical exponents characterizing
the whole critical region was necessary~\cite{sengupta08,viola08,divakaran08}.
A more general analysis of the problem in the context of the breakdown
of adiabaticity for gapless systems has been presented in Ref.~\cite{gritsev08}.
A numerical analysis of the raising of defects in a quenched spin chain model
exhibiting a BKT transition has been performed in Ref.~\cite{pellegrini08};
in that case defect formation is dominated by an isolated critical point,
so that a LZ treatment based on the finite-size closure of the dynamical gap
at that point is still possible.
On the other hand, one can also devise a KZ scaling argument, which relies
on the closing behavior of the gap as a function of the distance
from the critical point~\cite{polkov05,cucchietti07}.
In some circumstances this problem can be quite subtle, since it is possible
that the gap depends differently on the inverse size of the system
and on the parameter driving the transition, so that the two approaches give
different results:
this seems to be the case for the system considered in the present paper.
We are not aware of further quantitative studies of the dynamical defect
formation after an adiabatic crossing of the BKT transition line;
the major obstacle in understanding this type of dynamics raises
from the fact that here the scaling of defects is generally due to
multiple level crossings within the whole gapless phase.
We believe that this issue deserves further attention.
This is the aim of the present work.

The paper is organized as follows. In Sec.~\ref{sec:model} we introduce the model 
and recall the main features of its phase diagram.
In Sec.~\ref{sec:dynamics} we discuss the linear quenching scheme we adopt,
and define the excess energy of the system with respect
to the adiabatic limit: this quantity captures the essential physics of
the defect formation in the system. All the results of our work are 
concentrated in Sec.~\ref{sec:results}, while in Sec.~\ref{sec:concl} we 
draw our conclusions.

\section{The Model} \label{sec:model}

The Bose Hubbard (BH) model~\cite{fisher89},
well suited for describing interacting bosons in optical lattices~\cite{jaksch98},
is defined by the following Hamiltonian
\beq
   \Ham_{\rm BH} = - J \sum_i (a_i^\dagger a_{i+1} + {\rm h.c.})
   + \frac{U}{2} \sum_i n_i (n_i - 1) \, .
   \label{eq:BHmodel}
\eeq
Here $a_i^\dagger$ ($a_i$)  are the boson creation (annihilation)
operators on site $i$ (we assumed that the lattice is one-dimensional), and 
$n_i = a^\dagger_i a_i$ is the corresponding boson occupation number. 
The parameters $J$ and $U$ respectively denote the tunneling between
nearest neighbor lattice sites and the on-site interaction strength.
At integer fillings $1,2, \ldots$, when the ratio $t/U$ is gradually increased, 
the BH chain undergoes a QPT of the BKT type from a Mott insulating state, where 
bosons are localized in an incompressible phase, to a superfluid, with long range 
phase order.

The BH model in equation~\eref{eq:BHmodel} can be mapped into the effective spin-1 
Hamiltonian of equation~\eref{eq:SpinHam} in the limit of a large filling
and for small particle number fluctuations~\cite{altman02,huber07}.
When number fluctuations are not large it is possible to truncate the local Hilbert 
space to three  states with particle numbers $n_0$, $n_0 \pm 1$
($n_0$ being the average lattice filling per site). The reduced Hilbert space
of site $i$ can then be represented by three commuting bosons $t_{\alpha, i}$
($\alpha = -1,0,1$), which obey the constraint
$\sum_{\alpha=-1}^1 t^\dagger_{\alpha,i} t_{\alpha,i} = {\mathbb I}$.
In this way, the bosons of equation~\eref{eq:BHmodel} are represented by
$a^\dagger_i = \sqrt{n_0 + 1} \, t^\dagger_{1,i} t_{0,i}
+ \sqrt{n_0} \, t^\dagger_{0,i} t_{-1,i}$.
In the limit $n_0 \gg 1$ the effective Hamiltonian becomes 
the Hamiltonian of a one-dimensional spin-1 XY chain with single ion 
anisotropy~\cite{schulz86,denijs98},
\beq
  \label{eq:SpinHam}
  \Ham_{\rm spin} = -\frac{J_{\perp}}{2} \sum_i \left( S_{i}^{+} S_{i+1}^{-}
  +\text{h.c.} \right) + D \sum_i (S^{z}_{i})^{2} \, ,
\eeq
where $S_i^+ = \sqrt{2} (t^\dagger_{1,i} t_{0,i} + t^\dagger_{0,i} t_{-1,i})$,
$S_z^+ = t^\dagger_{1,i} t_{1,i} - t^\dagger_{-1,i} t_{-1,i}$
and with the identification
\begin{equation}
  \frac{Jn_0}{2} \to J_{\perp} \, , \qquad  U \to D
\end{equation}
(we chose to use the conventional notation for the spin-1 model).
In the previous equations $S^\alpha_i$ are spin-1 operators on site $i$
and $S^\pm_i = S^x_i \pm i S^y_i$;
$J_\perp$ and $D$ respectively characterize the nearest neighbor coupling
strength in the $xy$ plane and an uniaxial single-ion anisotropy along the
transverse $z$ direction.
This system is invariant under rotations around the $z$ axis, therefore the
total magnetization $S^{z}_{\rm tot} = \sum_{i} S^{z}_{i}$ is conserved. From 
now on all the quantities are expressed in units of the exchange coupling 
$J_{\perp} =1$.

\begin{figure}[!h]
  \begin{center}
    \includegraphics[scale=1.]{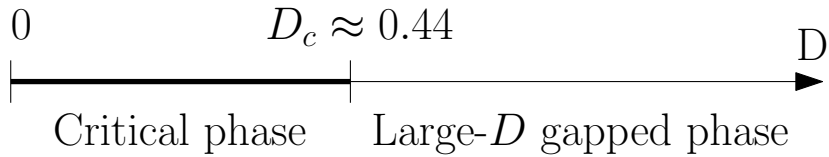}
    \caption{Phase diagram of the spin Hamiltonian~\eref{eq:SpinHam}
      for $J_\perp = 1$ and $D \geq 0$.}
    \label{fig:diag}
  \end{center}
\end{figure}

The phase diagram associated to the Hamiltonian in~\eref{eq:SpinHam} is sketched 
in figure~\ref{fig:diag}. For $D >0$ it consists in a large-$D$ phase for $D>D_{c}$, 
that is characterized by zero total magnetization (in the limit $D \to \infty$
each spin has zero magnetization), and a BKT transition line for $D\leq D_{c}$;
the critical point has been numerically estimated to be 
$D_{c}\simeq 0.44$~\cite{chen03,pires05,bologna}. In the rest of the paper we 
will only concentrate on the adiabatic dynamics of the Hamiltonian~\eref{eq:SpinHam}.

\section{Adiabatic dynamics} \label{sec:dynamics}

The adiabatic quench is realized by slowly changing the anisotropy 
parameter $D$ through the critical point $D_c$.
We suppose to vary $D$ linearly in time:
\begin{equation}
  D(t) = D_{\rm in} - \frac{t}{\tau}, \quad \text{with} \quad
  t \in \left[ 0, \tau (D_{\rm in} - D_{\rm fin}) \right] \,;
  \label{eq:quench}
\end{equation}
here $\tau$ is the quenching time scale $\tau$,
$D_{\rm in}$ and $D_{\rm fin}$ respectively denote the initial
and the final value of $D$.
In all the cases that will be analyzed we consider $D_{\rm in} > D_{c}$,
and suppose to initialize the system in its ground state;
on the other hand we take $D_{\rm fin} < D_c$, so that during the quench
the system crosses the BKT transition.
Since the initial ground state has zero total magnetization,
and this is conserved by the dynamics dictated by equation~\eref{eq:SpinHam},
only the excited states carrying zero magnetization will be accessible
throughout the quench.

In order to quantify the loss of adiabaticity of the system following
the quench, we study the behavior of the excess energy with respect
to the actual adiabatic ground state, after a proper rescaling:
\begin{equation}
  E_{\rm exc}(t) = \frac{ \bra{\psi(t)} \Ham(t) \ket{\psi(t)} -
  \bra{\psi_{GS}(t)} \Ham(t) \ket{\psi_{GS}(t)}} {\bra{\psi_{0}} \Ham(t) \ket{\psi_{0}}
  -\bra{\psi_{GS}(t)} \Ham(t) \ket{\psi_{GS}(t)}}
  \label{eq:excess}
\end{equation}
where $\ket{\psi_{0}}$ is the initial state of the system, that is the ground
state of Hamiltonian $\Ham(0)$; $\ket{\psi_{GS}(t)}$ is the instantaneous
ground state of $\Ham(t)$, and $\ket{\psi(t)}$ is the instantaneous wave function
of the system. Strictly speaking, the quantity $E_{\rm exc} (t)$ is not defined
at the initial time $t=0$, but one has $E_{\rm exc}(t \to 0^+) = 1$;
on the other hand at $t_f \equiv (D_{\rm in} - D_{\rm fin})/\tau$,
the excess energy gives, apart from a constant factor,
the final energy cost of defects in the system.
The final excess energy ranges from $E_{\rm exc}(t_f)=1$
(totally impulsive case) to $E_{\rm exc} (t_f) = 0$, for a fully
adiabatic evolution.

An exact solution for the spin model in equation~\eref{eq:SpinHam}
is not available, not even for the static case,
therefore one has to resort to numerical techniques.
In order to investigate both static properties and the dynamics after
the quench, we used the time-dependent Density Matrix Renormalization Group
(t-DMRG) algorithm with open boundary conditions~\cite{dmrg}.
For the dynamics at small sizes $L \leq 10$, we checked
our t-DMRG results with an exact numerical algorithm which does not
truncate the Hilbert space of the system.
For static computations we were able to reach sizes of $L=200$,
while for dynamics simulations we considered systems of up to $L=80$ sites.
The time evolution has been performed with a second order Trotter
expansion of $\Ham(t)$; in most simulations we chose a discretization time
step $\delta t=10^{-3}$, while the truncated Hilbert space dimension has been
set up to $m=200$.

\section{Results} \label{sec:results}

In this section we describe our results for the adiabatic dynamics of the spin-1
Hamiltonian. 
We first analyze the behavior of the excitation gaps which are relevant
for the quenched dynamics. Then we focus on the dynamics, and discuss the behavior
of the excess energy~\eref{eq:excess} as a function of
the quenching rate $\tau$. We first consider the slow-quench
region for small system sizes and then concentrate on the scaling regime for larger 
sizes.

\subsection{Dynamical gap}

A great deal of understanding on the adiabatic dynamics derives from the 
knowledge of the finite size scaling of the first excitations gaps.
As stated before, since the dynamics of the system conserves the total $z$
magnetization, if we suppose to start from the zero-magnetization ground
state, only excited states with $S^z_{\rm tot}=0$ will be involved during
the dynamics.
Therefore, the {\it dynamical gap} is defined as the first relevant gap
for the dynamics, that is the energy difference between the ground state
and the first excited state compatible with the integrals of motion.

\begin{figure}[!t]
  \begin{center}
    \includegraphics[scale=0.5]{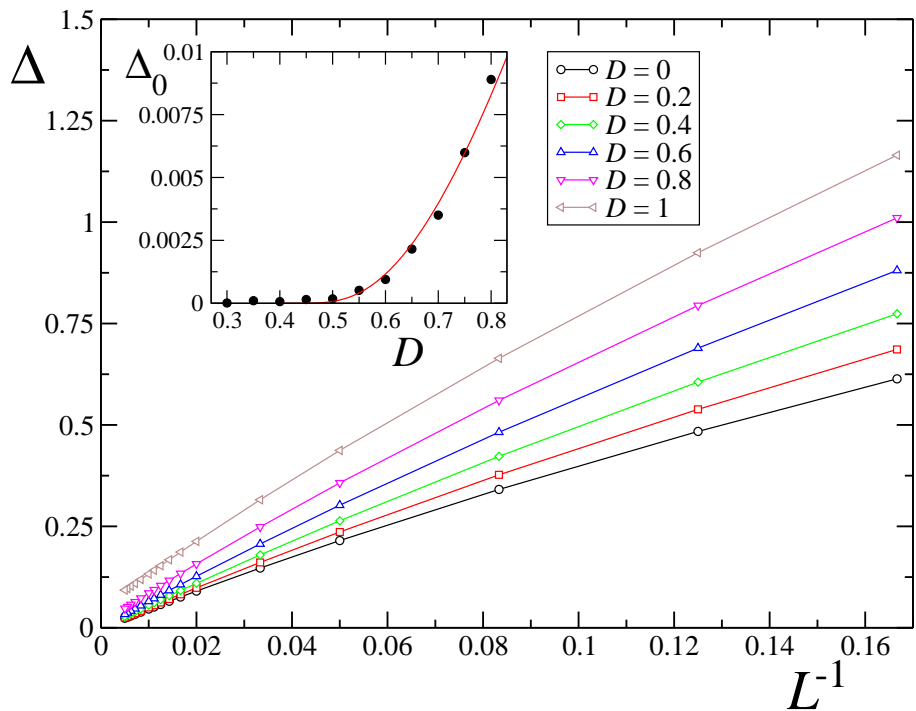}
    \caption{Ground state excitation energy $\Delta$ in the zero magnetization
      sector as a function of the inverse system size $L^{-1}$
      (we show data ranging from $L=6$ to $L=200$).
      The various curves are for different values of the single-ion anisotropy $D$.
      In the inset we plot the asymptotic value $\Delta_0$ in the thermodynamic
      limit, as extracted from a quadratic fit of the data in main panel
      for $L \geq 50$ (black circles); the red line displays a fit
      $\Delta_0 \propto \exp(-c/\sqrt{D-0.44})$ of data with $c\approx 2.977$.}
    \label{fig:gap_size}
  \end{center}
\end{figure}

As shown in figure~\ref{fig:gap_size}, in the critical region $D<D_c$
the dynamical gap $\Delta$ scales approximately linearly as a function
of the inverse system size $L^{-1}$. The same behavior also holds for relatively
small values of $D-D_c$ within the gapped phase, as those considered
in figure~\ref{fig:gap_size}, so that the correlation length is still larger
than the size of the system, and a quasi-critical regime is found~\cite{bologna}.
On the other hand, we numerically checked that the leading term in finite-size
corrections scale as $L^{-2}$ for $D \gtrsim 2$, when the system is far from
criticality.
We extrapolated the value of $\Delta_0=\Delta(L\to \infty)$ in the thermodynamic limit
by performing a fit of numerical data for $L \geq 50$ which includes
both the leading linear behavior and smaller quadratic corrections.
Results are plotted in the inset of figure~\ref{fig:gap_size}, as a function of $D$.
According to the phase diagram of the system,
which predicts a closure of the gap for $D < D_c \approx 0.44$,
the asymptotic value of the gap is found to be constant
and equal to zero for $0 \leq D \lesssim 0.45$
(up to values $\sim 2 \times 10^{-4}$), while it suddenly raises up
as $\Delta_0 \sim \exp(-c/\sqrt{D-D_c})$ in the gapped phase close to criticality.
Thus the dynamical gap closes analogously to the gap between the ground state
and the first excited state with unconstrained magnetization,
which is called \emph{thermodynamical gap}, in a BKT transition~\cite{chaikin95}.

\begin{figure}[!t]
  \begin{center}
    \includegraphics[scale=0.5]{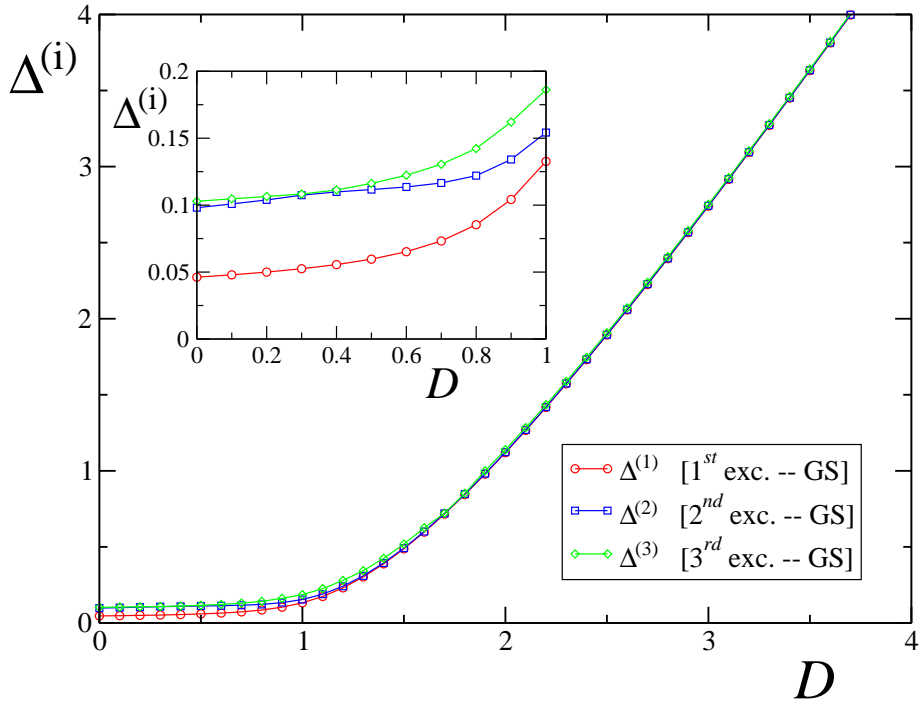}
    \caption{Excitation energy $\Delta^{(i)}$ of the three lowest excited
      dynamical levels for $L=100$ spins in the subspace $S^{z}_{\rm tot} = 0$,
      as a function of $D$; the first excitation energy coincides with
      the dynamical gap: $\Delta^{(1)} \equiv \Delta$.
      The inset shows a zoom for $0 \leq D \leq1$ of the same plot.}
    \label{fig:gap100art}
  \end{center}
\end{figure}

The excitation energies of the first three dynamical excited levels in
the subspace of zero magnetization and for a system of $L = 100$ sites
are displayed in figure~\ref{fig:gap100art},
as a function of the anisotropy $D$. In the large-$D$ phase the dynamical
gap $\Delta^{(1)} \equiv \Delta$ is well above the zero;
when decreasing $D$ it closes approximately linearly until $D \sim 2$,
then it continues closing as far as it approaches a region
for $D\lesssim 0.5$, where it becomes almost constant and very small,
as shown in the inset.

We point out that this type of behavior is quite different from the
scenario elucidated in the spin-1/2 Heisenberg model of Ref.~\cite{pellegrini08}.
In that case two types of quenches involving the
antiferromagnetic BKT isotropic point were considered.
While in the second quenching scheme the system started from the critical
region and advanced in the opposite direction with respect to our case, the first quench
started from the antiferromagnetic region and crossed both the BKT
point and the ferromagnetic isotropic point.
Remarkably, the excess energy was found to be essentially characterized
by the features of the ferromagnetic critical point, where the gap closes
faster than in all the other points along the critical line.
Therefore it was possible to identify a dominant critical point which
allowed for the applicability of a LZ scaling argument in determining
the defect density.
On the other hand, our quench involves the BKT transition line close
to the antiferromagnetic isotropic point, and there are no dominant critical points,
thus leading to a more complex scenario, as explained in the following.

\subsection{Oscillations in the excess energy for slow quenches}

\begin{figure}[!t]
  \begin{center}
    \includegraphics[scale=0.5]{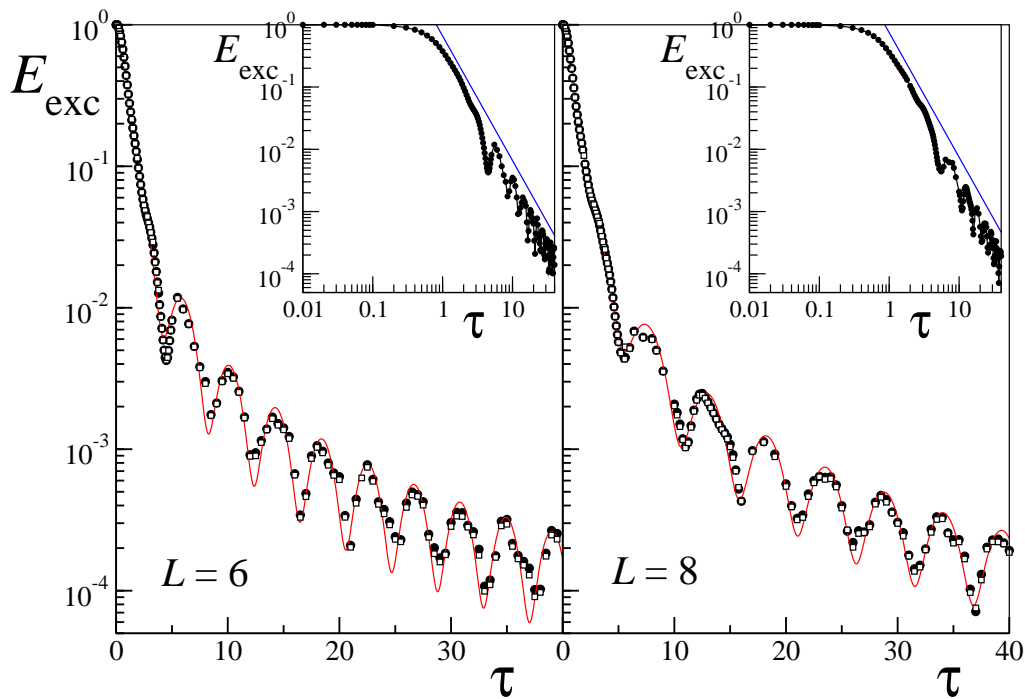}
    \caption{Final excess energy after an adiabatic quench of $D$ from
      $D_{\rm in}=1$ to $D_{\rm fin}=0$, as a function of
      the quench velocity $\tau$.
      Filled circles denote t-DMRG data, empty squares are obtained with
      exact diagonalization, while the continuous line
      is a numerical fit with the formula predicted by a LZ model for
      finite initial and final coupling times.
      The left panel shows data for $L=6$ sites, while the right one
      is for $L=8$. 
      The insets show the same data in a log-log scale
      (straight blue lines denote a $\sim \tau^{-2}$ behavior).
      Note the smaller frequency of the oscillations for $L=8$.}
    \label{fig:Vitanov_D1-0}
  \end{center}
\end{figure}

Let us first consider systems of small sizes, as shown
in figure~\ref{fig:Vitanov_D1-0} for $L=6$ and $L=8$ sites.
We have evaluated the excess energy both with the t-DMRG algorithm
(filled circles), and with an exact diagonalization which does not truncate
the system's Hilbert space (empty squares). As the figure shows,
data agree well.

On increasing the rate $\tau$, we can recognize
two different regimes. For very small values of $\tau$
the excess energy is close to its maximum, and the dependence
on the size and on $\tau$ is very small.
These points correspond to very fast quenches, where the system
dynamics is strongly non-adiabatic and the initial state is substantially frozen.
As a consequence, the state after the quench is found to be in a superposition
of many excited states of the final Hamiltonian.
A second region is characterized by a dominant power-law decay,
according to $E_{\rm exc} \sim \tau^{-2}$ (see the straight lines in the
two insets of figure~\ref{fig:Vitanov_D1-0}), that is superimposed to an oscillatory behavior.
This can be explained within a LZ approximation:
for small values of $D$ and $L$ the gap is large and proportional to $1/L$,
therefore at very small sizes only the ground state and the first excited state
participate to the evolution of the system, while all the other excited states
are not accessible.
The power-law decay, as well as the oscillations naturally arise
when effects of finite duration time are taken into account~\cite{vitanov}. 
Following the closing of the gap, the frequency of the oscillations decreases
at increasing sizes, as it can be seen in the figure.
The red curve displays a fit of numerical data obtained by an effective
LZ model in which the initial coupling time $t_i <0$ is finite,
and the final time is $t_f = 0$
(see~\ref{sec:LZfinite} for details on the fitting formula).

\begin{figure}[!t]
  \begin{center}
    \includegraphics[scale=0.5]{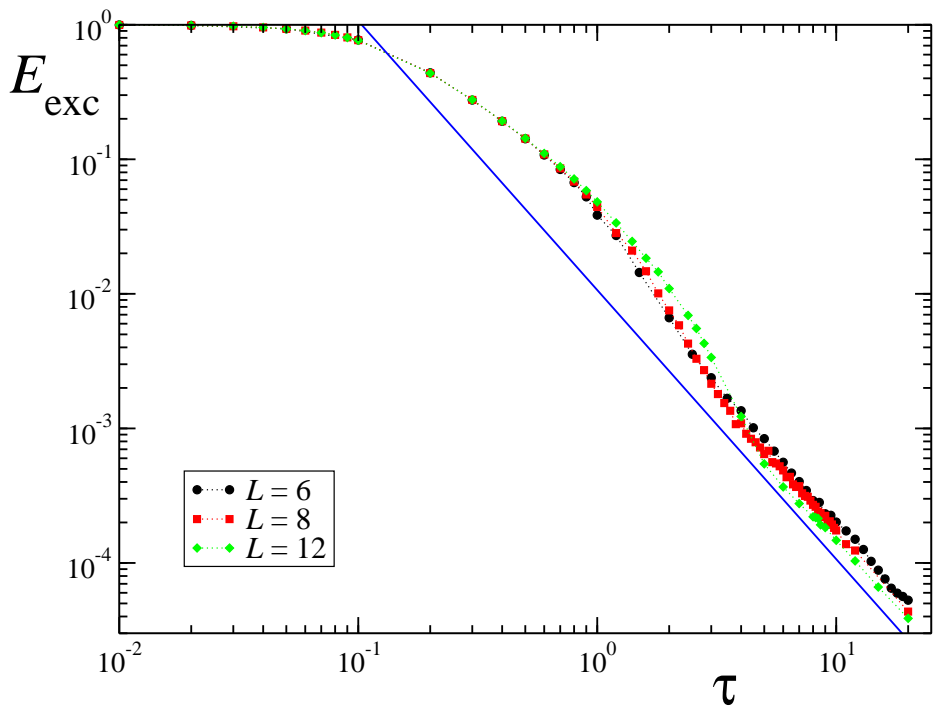}
    \caption{Excess energy after an adiabatic quench of $D$ from
      $D_{\rm in}=4$ to $D_{\rm fin}=0$, as a function of $\tau$.
      The various data are for different system sizes
      $L=6$ (black circles), 8 (red squares), 12 (green diamonds).
      The straight blue line indicates a behavior
      $E_{\rm exc} \sim \tau^{-2}$ and is plotted as a guideline.}
    \label{fig:EnRes_4-0}
  \end{center}
\end{figure}

The oscillatory behavior can be drastically suppressed starting
from a larger value of $D_{\rm in}$, which corresponds, in the LZ model,
to decreasing the initial coupling time $t_i$; for $t_{f} \leq 0$ and
in the limit of $t_i \to - \infty$ the oscillations disappear
and a pure power-law $\sim \tau^{-2}$ decay survives~\cite{vitanov}.
This is seen to emerge from numerical data of figure~\ref{fig:EnRes_4-0},
where we started quenching from $D_{\rm in} = 4$.
Notice also the substantial independence of $E_{\rm exc}$
on the system size in the fast quenching limit.

\subsection{Scaling regime}

The analysis of the effects of the quantum phase transitions on the adiabatic
quench dynamics demands sufficiently large system sizes.
We now concentrate on this aspect and study the excess energy as a function
of $\tau$ for considerably larger values of $L$.
Due to the increasing computational difficulty in simulating large systems,
we restrict ourselves to quenching schemes in which $D_{\rm in}=1$.

\begin{figure}[!t]
  \begin{center}
    \includegraphics[scale=0.5]{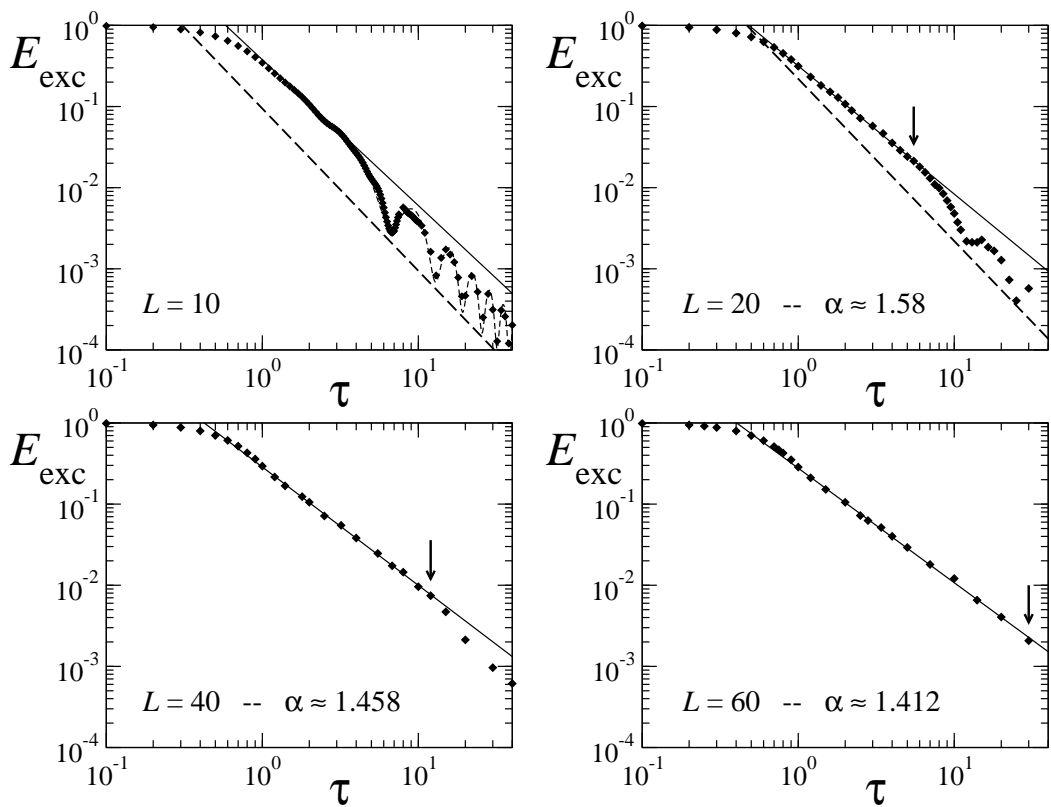}
    \caption{Final excess energy as a function of the quench rate $\tau$.
      The various panels stand for different system sizes.
      Symbols denote numerical t-DMRG data, while the straight line
      is a power-law fit that has been performed for $\tau < \tau^*$
      ($\tau^*$ is indicated by the vertical arrow).
      The two straight dashed lines in the upper panels denote a $\tau^{-2}$
      behavior, and are plotted as guidelines.
      The oscillating dashed line at $L=10$ is a fit of data with
      big $\tau$, according to the LZ model for a finite coupling duration.
      The values of $\alpha$ corresponding to the power-law fits are
      quoted in each panel.
      Here we set $D_{\rm in} = 1$ and $D_{\rm fin} = 0$.}
    \label{fig:En_Res}
  \end{center}
\end{figure}

In figure~\ref{fig:En_Res} we plot the final excess
energy of the system after a quench from $D_{\rm in}=1$ to
$D_{\rm fin}=0$ of time duration $\tau$.
Starting from fast quenches and going towards slower ones,
we can now distinguish three different regimes: the first strongly non-adiabatic
regime at small $\tau$ is analogous to the one previously discussed
for small sizes. In the opposite limit of very slow quenches $\tau \gg 1$,
we also recover the power-law $\tau^{-2}$ behavior superimposed to
oscillations coming from an effective LZ description with finite
coupling duration.
Most interestingly, in between these two opposite situations,
a characteristic power-law regime emerges, where:
\beq
   E_{\rm exc} \sim \tau^{-\alpha} \, \quad {\rm with} \;\; \alpha \in [1,2] \, .
   \label{eq:scaling}
\eeq
This is dominated by transitions to the lowest dynamically accessible gap,
and it is crucially affected by the critical properties of the system.
The crossover time $\tau^{*}$ at which this regime ends typically increases
with the size, as it can be qualitatively seen from the figure
(arrows denote a rough estimate of $\tau^{*}$ for the different sizes),
and diverges in the thermodynamic limit; unfortunately we were not able to analyze
the scaling with $L$, because of the intrinsic difficulty in estimating
the ending point of the $\tau^{-\alpha}$ behavior.
Nonetheless, even at asymptotically small quenching velocities,
for very large sizes the scaling of defects~\eref{eq:scaling} ruled
by criticality persists, thus meaning that the system dynamics
cannot be strictly adiabatic.

The scaling of the decay rate $\alpha$ with the size has been analyzed
numerically, for data corresponding to $L$ ranging from $10$ to $60$ spins;
at $L < 10$ this regime was not identifiable.
Some representative cases are shown in figure~\ref{fig:En_Res}, where
each of the four panels stands for a given system size, while
straight continuous lines indicate the best power-law fits of the scaling regions.
In the case of $L=10$ sites (upper left panel), we cannot give a
reliable estimate of $\alpha$, since the width of the scaling region is narrow
and the fit is very sensitive to its actual starting and ending points.
The straight line in the plot corresponds to $\alpha \approx 1.798$ and has
been obtained from a power-law fit of numerical data from $\tau = 1$ to $\tau^* = 3$.
As one can see, this is hardly distinguishable from the $\tau^{-2}$ power-law behavior
of the slow-quench regime (straight dashed line), thus meaning that
the existence of the scaling region itself is here in doubt.
This is not the case for the other panels, where a power-law behavior of the
type in equation~\eref{eq:scaling} is clearly visible. Namely,
we fitted our data until the $\tau^*$ value, that is labeled
in figure~\ref{fig:En_Res} by a vertical arrow:
as we could expect, the size of the scaling region increases with $L$.

\begin{figure}[!t]
  \begin{center}
    \includegraphics[scale=0.5]{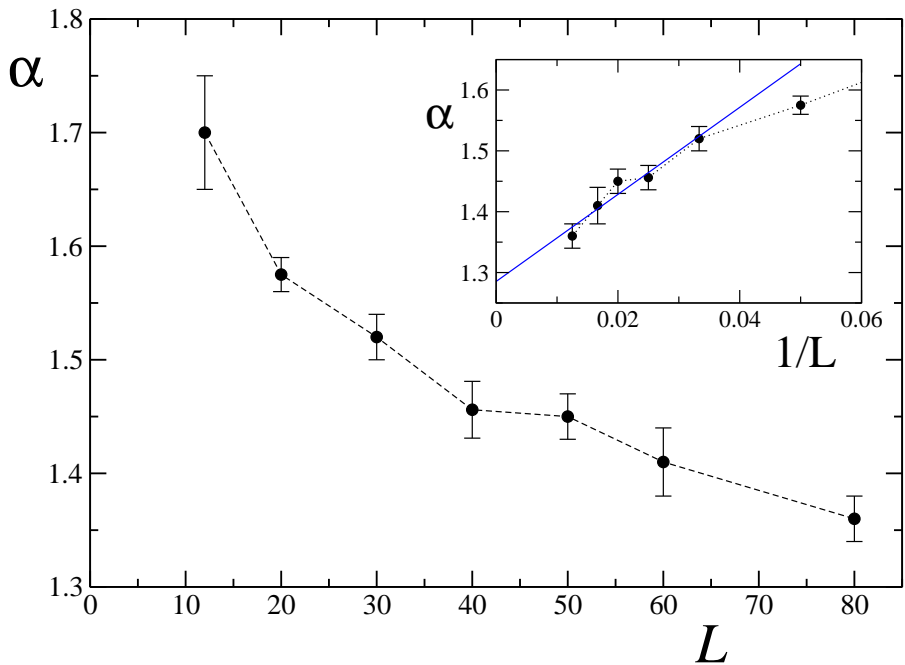}
    \caption{Power-law decay rate $\alpha$ in the intermediate scaling region
      for the excess energy, as a function of the system size $L$.
      The quench is performed from $D_{\rm in} = 1$ to $D_{\rm fin} = 0$.
      In the inset we plot the same quantity as a function of $1/L$.
      The blue line is a linear fit of data with $L \geq 30$, and
      predicts an asymptotic value of $\alpha_\infty \approx 1.28$ in the
      thermodynamic limit.}
    \label{fig:Alpha_Tau}
  \end{center}
\end{figure}

Summarizing the results obtained for the various sizes,
in figure~\ref{fig:Alpha_Tau} we report the behavior of $\alpha$ as a function
of $L$ (in the inset we plot the same data with $1/L$ on the $x$-axis).
The uncertainty affecting the value of $\alpha$ extracted from the power-law
fits of numerical t-DMRG data is mostly due to the inaccurate knowledge
of the extremes of the scaling region.
For each value of $L$, we identified a trial power-law region and then computed
several values of $\alpha$ by progressively sweeping out the points from that
region, starting from the borders.
We then evaluated error bars, that are displayed in the plot,
by performing a statistical analysis of the values of $\alpha$ thus obtained.
In order to give an estimate of the power-law decay rate in the
thermodynamic limit, we supposed that, at large $L$, $\alpha$ scales
inversely proportional with the system size.
In this way, performing a linear fit of data with $L \geq 30$, we extracted
the asymptotic value $\alpha_\infty \approx 1.28$
in the thermodynamic limit (see straight blue line in the inset).

We would like to stress that, in this context, our numerical results
seem not to find a straightforward explanation with LZ or KZ arguments.
One could, for example, try to follow a standard LZ argument,
that is based on the assumptions that the dynamical gap scales linearly
with the inverse size, nearby and inside the critical region,
and that the adiabaticity loss is essentially due
to the presence of a dominant critical point, where the gap closes faster
than elsewhere~\cite{zurek05,pellegrini08}.
In the LZ approximation, the probability of exciting the ground state
is a global function of the product $\tau \Delta_{\rm m}^2$, where $\Delta_{\rm m}$
is the minimum gap achieved by the system during the quench.
Assuming a critical scaling of the gap $\Delta_{\rm m} \sim L^{-1}$,
as shown in figure~\ref{fig:gap_size}, the density of defects
can be estimated by evaluating the typical length $L_\varepsilon$ of a
defect-free region, once the probability for this to occur is $\varepsilon$.
As a consequence, this would give
$E_{\rm exc} \sim 1/L_\varepsilon \propto \tau^{-1/2}$, 
exactly as in the Ising case, in contrast with numerical evidence.
On the other hand, a scaling argument based on the KZ mechanism
can be adopted~\cite{zurek05,polkov05}; this relies
on the fact that the dynamical gap $\Delta_0$ seems to close with $D-D_{C}$
as the thermodynamical gap in a BKT transition, that is,
it depends on the anisotropy parameter
as $\Delta_0 \propto \exp(-c/\sqrt{D-D_{C}})$~\cite{chaikin95},
so that the critical exponent for the correlation length $\nu$ diverges.
In this case, the KZ scaling argument predicts a power-law scaling exponent
$\alpha = (d+z)\nu/(z\nu+1)$, $d$ being the dimension of the system
and $z, \nu$ critical exponents; in our specific case $d=z=1, \nu \to \infty$
thus leading to $\alpha = 2$ (plus some logarithmic
corrections)~\cite{polkov05,cucchietti07}.
This again contrasts with our numerical evidence, thus revealing that
the presence of a critical line in which the gap closes always in the same way
seems to indicate that all the low-lying excitation spectrum becomes
necessary to predict the actual behavior.
We notice that the two above mentioned different dependencies of the dynamical
gap on the inverse size, like $\sim1/L$ for $L\to\infty$, and on the distance
from the critical point, as $\sim e^{-c/\sqrt{D-D_{c}}}$,
are confirmed quite precisely by our data.

\begin{figure}[!t]
  \begin{center}
    \includegraphics[scale=0.5]{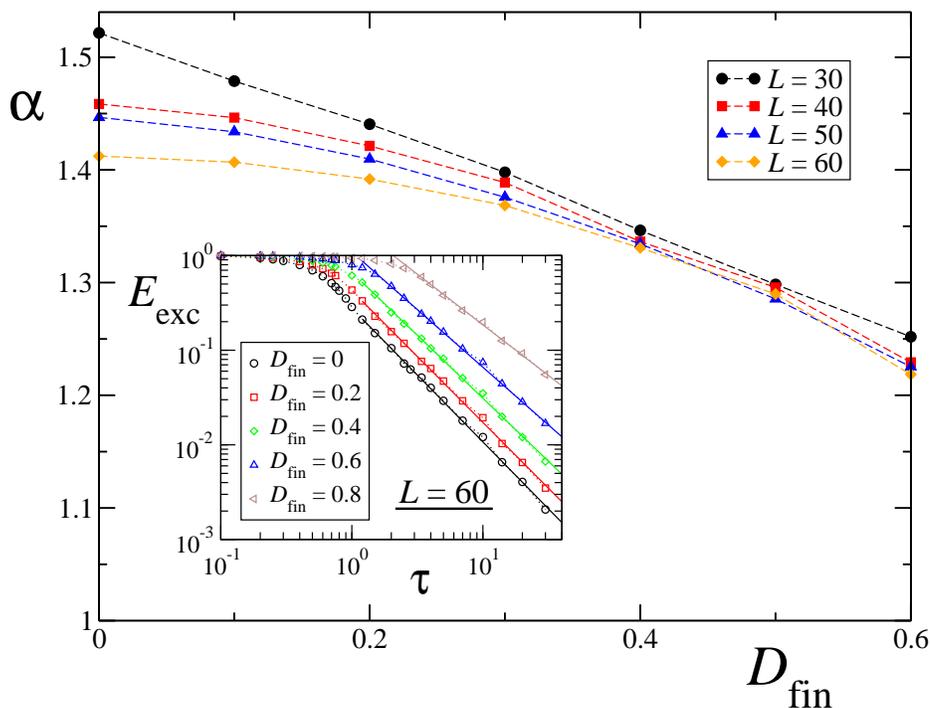}
    \caption{Power-law decay rate $\alpha$ as a function
      of the ending value for the quench $D_{\rm fin}< D_{\rm c}$
      and for fixed $D_{\rm in} = 1$. Data are for different system sizes,
      as explained in the caption.
      In the inset we display the excess energy for a quench ending
      at various $D_{\rm fin}$, and for a system size $L=60$.
      straight lines are power-law fits in the scaling regime.}
    \label{fig:EnRes_Dvar}
  \end{center}
\end{figure}

We checked the dependence of $\alpha$ on the final point of the quench: 
in figure~\ref{fig:EnRes_Dvar} we varied the ending point $D_{\rm fin}$, while keeping
$D_{\rm in}$ and the system size fixed (explicit data for the excess energy
$E_{\rm exc}$ as a function of $\tau$ are presented in the
inset, at $L=60$). For values of $D_{\rm fin}$ outside the critical region we
find that $\alpha$ depends on $D_{\rm fin}$ at finite sizes. Nevertheless, the range of the scaling 
region shrinks with $L$ and eventually disappears in the thermodynamic limit, so we argue 
that the dependence of $\alpha$ on $D_{\rm fin}>D_{\rm c}$ 
should be entirely due to finite size effects. For $D_{\rm fin}<D_{\rm c}$ we observe that
the dependence of $\alpha$ on $D_{\rm fin}$ weakens as the system size is increased. In this 
case the scaling region is valid until a quench rate $\tau^* \stackrel{L \to \infty}{\longrightarrow} + \infty$;
the power-law decay rate tends to a value that is independent of $D_{\rm fin} < D_c$ and
has been extrapolated from numerical data of figure~\ref{fig:Alpha_Tau}
to be $\alpha_\infty \approx 1.28$.

\section{Conclusions} \label{sec:concl}

In this work we have analyzed the quenched dynamics of a quantum anisotropic
spin-1 XY chain, when it crosses a Berezinskii-Kosterlitz-Thouless quantum phase transition.
The quench has been performed on the uniaxial single-spin anisotropy,
and has been chosen to vary linearly in time with a given velocity.
We focused on the residual excess energy of the system after the quench,
and studied its dependence on the velocity of the quench.
For very slow quenches and finite system sizes we were able to describe
the properties of the system in terms of an effective Landau Zener model,
where the system can only get excited to its first excited state.
Most interestingly, we pointed out the emergence of an intermediate region
where the excess energy drops as a power-law with the quench rate,
and exhibits a non trivial scaling behavior.
At least for the finite sizes considered here, the decay rate depends on the size
of the crossed critical region, and cannot be explained in terms
of usual scaling arguments, such as the
standard Kibble-Zurek mechanism and its generalization to critical
surfaces~\cite{pellegrini08,sengupta08,viola08,divakaran08}.
In the thermodynamic limit the system obeys a non-trivial scaling behavior
$E_{\rm exc}\sim \tau^{-\alpha}$, 
with $1<\alpha<2$, even when $\tau\to \infty$ (i.e., for very slow quenches).

\ack

We acknowledge fruitful discussions with Tommaso Caneva, Vladimir Gritsev,
Simone Montangero, Franco Pellegrini, Anatoli Polkovnikov and Alessandro Silva,
and financial support of EU through the Integrated Project EUROSQIP.
The Numerical t-DMRG simulations in this work have been performed
using the code released within the ``Powder with Power''
project, available at: \texttt{http://www.dmrg.it}.

\appendix

\section{Landau-Zener model for finite coupling times} \label{sec:LZfinite}

The Landau-Zener (LZ) model consists in a two level system describing
an avoided level crossing: two energy levels moving in time are widely
separated at first, then they approach each other with time, and finally
part away again~\cite{landau,zener}.
When the two levels are well separated, each eigenstate preserves an
individual character; on the other hand, when levels are close together,
they mix due to their interaction.
The Hamiltonian is given by:
\beq \label{eq:lz}
   \Ham_{LZ} = \left( \begin{array}{cc}
   -\Delta(t) & \Omega \\ \Omega & \Delta(t)
   \end{array} \right),
\eeq
with a detuning $\Delta(t)=\beta^{2}t$ (where $\beta^2 >0$),
and a time independent coupling $\Omega$ that, in the original LZ
model is supposed to last from $t_i = - \infty$
to $t_f = + \infty$~\cite{landau,zener}.
Here we review the general case where the coupling is turned on
at $t_i$ and off at $t_f$~\cite{vitanov}.
The equation~\eref{eq:lz} is written in the basis
of the two eigenstates of the Hamiltonian in absence of interaction.

The probability amplitudes $\vec{C}(t_i) = [ C_1(t_i), \, C_2(t_i) ]^T$
for the two levels at the beginning are connected to the ones at the
final time $t_f$ by the unitary evolution matrix $U(t_f,t_i)$, so that:
$\vec{C}(t_f) = U(t_f,t_i) \, \vec{C}(t_i)$. Their elements are given by:
\begin{eqnarray} \label{eq:Uvit1}
   U_{11}(T_{f},T_{i}) & =\frac{\Gamma(1-\frac{1}{2} i \omega^2)}{\sqrt{2 \pi}}
   [ D_{i\omega^{2}/2}(T_{f}\sqrt2 e^{-i\pi/4}) \\
  & \times D_{-1+i\omega^{2}/2}(T_{i}\sqrt2 e^{i3\pi/4})+D_{i\omega^{2}/2}(T_{f}\sqrt2 e^{i3\pi/4}) \nonumber\\
  & \times D_{-1+i\omega^{2}/2}(T_{i}\sqrt2 e^{-i\pi/4})]\nonumber \, ,
\end{eqnarray}
\begin{eqnarray} \label{eq:Uvit2}
   U_{12}(T_{f},T_{i}) & =\frac{\Gamma(1-\frac{1}{2} i \omega^2)}{\omega \sqrt{\pi}} e^{i \pi /4}
   [ -D_{i\omega^{2}/2}(T_{f}\sqrt2 e^{-i\pi/4}) \\
   &\times D_{i\omega^{2}/2}(T_{i}\sqrt2 e^{i3\pi/4})\nonumber \\
   & + D_{i\omega^{2}/2}(T_{f}\sqrt2 e^{i3\pi/4}) \, D_{i\omega^{2}/2}(T_{i}\sqrt2 e^{-i\pi/4})]\nonumber \, ,
\end{eqnarray}
where we have introduced the rescaled time $T = \beta t$ and the scaled
dimensionless coupling strength $\omega = \Omega / \beta$,
while $D_{\nu}(z)$ denote the parabolic cylinder functions.
If we suppose that the system is initialized in its ground state, i.e.,
$C_1(t_i) = 1, \; C_2(t_i) = 0$, the transition probability to the excited
state at the final time is given by
$P^{(d)} (t_f, t_i) = \vert U_{21} (t_f, t_i) \vert^2$.

This is related to the relevant adiabatic basis, that is the basis
of the instantaneous system eigenstates, by a unitary transformation.
If $\vec{A}(t) = [ A_1(t), \, A_2(t) ]^T$ are the probability amplitudes
for the two levels in the adiabatic basis, then
$\vec{A}(t) = {\bf R}(t) \, \vec{C}(t)$, where
${\bf R}(t)$ is the rotation matrix
\beq
    {\bf R}(T) = \left( \begin{array}{cc}
    \cos\vartheta(t) & -\sin\vartheta(t)\\ \sin\vartheta(t) & \cos\vartheta(t)
    \end{array} \right),
\eeq
with $\tan[ 2 \vartheta(t)] = \Omega(t) / \Delta(t)$.
Therefore, the evolution matrix in the adiabatic representation is
given by $U_a(t_f,t_i) = {\bf R}^T (t_f) \, U(t_f, t_i) \, {\bf R} (t_i)$,
and the adiabatic-following solution for the transition probability
is $P^{(a)} (t_f , t_i) = \vert U^{(a)}_{21} (t_f, t_i) \vert^2$.

For the original LZ model, where the coupling is supposed to last
from $t_i \rightarrow -\infty$ to $t_f \rightarrow +\infty$,
the expression for the excitation probability at the end of the quench
in the adiabatic basis simplifies to an exponential form:
\beq \label{eq:pinfty}
P^{(a)}(+\infty,-\infty)=e^{-\pi \omega^{2}} \, .
\eeq
In the case of a \emph{finite} coupling duration, that ends before or exactly
at the crossing (i.e., $t_f \leq 0$),
we have a much involved expression, which predicts a leading power-law behavior
$P^{(a)} \sim \tau^{-2}$ superimposed to an oscillating behavior.
Eventually oscillations are damped for long lasting couplings:
in the limiting case where the quench ends at the critical point
and is infinite lasting ($t_i = - \infty, \, t_f = 0$),
the probability is given by
\beq \label{eq:p-infty0}
P^{(a)}(0,-\infty) = \frac{1}{16 \, \omega^{4}} \sim \frac{1}{\tau^2}\, .
\eeq
The scaling with the quench velocity $\tau$ follows from the fact that
the times $t \propto \tau$, while $\beta^2 \propto 1/\tau$
(this implies that $\omega \propto \sqrt{\tau}$).

We used the explicit formula for the adiabatic transition probability
$P^{(a)}(0,t_i<0)$ in order to fit t-DMRG data for the excess energy
of our system in the regime of large $\tau$, where defects still do not
form and the quench dynamics can be considered adiabatic.
While it is clear that $t_{i}<0$ in our case, it is not obvious a priori
whether $t_{f}<0$ or $t_{f}=0$, since LZ relies on the assumption that there is
only one point of closest approach of the energy levels; on the contrary,
in our model we have a whole critical line.
We actually chose $t_{f}=0$ and used $t_{i}<0$ as a fitting parameter,
having not a rigorous criterion at our disposal, but following the
qualitative picture suggested from figure~\ref{fig:gap100art}:
the gap closes monotonically during the quench,
reaching the minimum at $D_{\rm fin}$.
The red curves in figure~\ref{fig:Vitanov_D1-0} have been obtained by fitting
numerical data with the theoretical prediction given by $P^{(a)}$;
we admitted a global rescaling prefactor $\phi$ and imposed the following
constraints: $T_i = - T_0 \sqrt{\tau}$, $T_f = 0$, $\omega = \omega_0 \sqrt{\tau}$.
The fitting parameters are $T_0, \, \omega_0, \, \phi$.
For the left panel ($N=6$) we chose
$T_0 \approx 0.8, \, \omega_0 = 0.84, \, \phi = 1.72$, while for
the right one ($N=8$) $T_0 \approx 0.735, \, \omega_0 = 0.7, \, \phi = 0.9$.

\end{document}